\documentclass[aps,twocolumn,prb,amsmath,amssymb,superscriptaddress,floats,nobibnotes,nofootinbib]{revtex4}

\usepackage{CJK}                     
\usepackage{amsmath}	
\usepackage{gensymb}
\usepackage[colorlinks=true,linkcolor=blue,urlcolor=blue,citecolor=blue]{hyperref}
\hypersetup{colorlinks=true}
\usepackage{upgreek}
\usepackage{graphics}
\usepackage{hyperref}
\usepackage{epsfig}
\usepackage{color}
\usepackage{bm}
\usepackage{float}
\usepackage{ulem} 
\usepackage{dsfont}                 
\usepackage{wasysym}
\usepackage{multirow}
\usepackage{amssymb}

\usepackage{blindtext}

\usepackage{accents}

\usepackage{graphicx}

\graphicspath{{./}{./plots/}}

\usepackage{url}
\usepackage{multirow}

\newcommand{\upperRomannumeral}[1]{\uppercase\expandafter{\romannumeral#1}}

\begin{document}
	
	\begin{CJK*}{GBK}{song}

		\title{Strong enhancement of $d$-wave superconductivity in an extended checkerboard Hubbard ladder}		
		
		\author{Xichen Huang}
		\affiliation{Lanzhou Center for Theoretical Physics, Lanzhou University, Lanzhou 730000, China}
		\affiliation{Key Laboratory of Quantum Theory and Applications of MoE, Lanzhou University, Lanzhou 730000, China}
		\affiliation{Key Laboratory of Theoretical Physics of Gansu Province$\&$Gansu Provincial Research Center for Basic Disciplines of Quantum Physics, Lanzhou University, Lanzhou 730000, China}
		
		\author{Saisai He}
		\affiliation{Lanzhou Center for Theoretical Physics, Lanzhou University, Lanzhou 730000, China}
		\affiliation{Key Laboratory of Quantum Theory and Applications of MoE, Lanzhou University, Lanzhou 730000, China}
		\affiliation{Key Laboratory of Theoretical Physics of Gansu Province$\&$Gansu Provincial Research Center for Basic Disciplines of Quantum Physics, Lanzhou University, Lanzhou 730000, China}
		
		\author{Jize Zhao}
		\email[]{zhaojz@lzu.edu.cn}
		\affiliation{Lanzhou Center for Theoretical Physics, Lanzhou University, Lanzhou 730000, China}
		\affiliation{Key Laboratory of Quantum Theory and Applications of MoE, Lanzhou University, Lanzhou 730000, China}
		\affiliation{Key Laboratory of Theoretical Physics of Gansu Province$\&$Gansu Provincial Research Center for Basic Disciplines of Quantum Physics, Lanzhou University, Lanzhou 730000, China}
		
		\author{Zhong-Bing Huang}
		\email[]{huangzb@hubu.edu.cn}
		\affiliation{School of Physics, Hubei University, Wuhan 430062, China}
		
		\date{\today}
		
		\begin{abstract}
			{
				By employing the density-matrix renormalization group method, we study an extended checkerboard Hubbard model on the two-leg ladder, 
				which includes an intraplaquette nearest-neighbour attraction $V$. The simulated results show that $V$ 
				plays a significant role in enhancing the $d$-wave superconductivity when the electron density is close to half-filling. 
				In the homogeneous case $t'=t$ ($t$ and $t'$ are the intraplaquette and interplaquette hopping integrals), large critical $|V_{c}|$ 
				is required to induce the superconducting ground state. With decreasing $t'$, $|V_{c}|$ is substantially diminished and 
				the pair state has a nearly $C_4$ symmetry. In the extremely inhomogeneous case $t'<0.2t$, the system transits 
				to the $d$-wave superconducting phase at $V\sim -0.3t$ and $V\sim -0.4t$ for $U=8t$ and $U=12t$, respectively, accompanying with 
				a shift of spin and single-particle excitations from gapless to gapped type.}
		\end{abstract}
		
		\maketitle
		
		\section{Introduction}
		Although lots of experiments have confirmed anisotropic $d$-wave superconductivity in the high-$T_c$ cuprates over the past decades\cite{HF_Feng1999,CC_Tsuei2000,SH_Pan2001,ZX_Shen1993}, 
		the associated microscopic mechanism remains illusive and continues to be the research focus in condensed matter physics\cite{Dagotto1994challenge}.    
		The underlying difficulty may arise partially from strong electron correlation effect in the high-$T_c$ cuprates, and thus the weak-coupling 
		approximation, which is central in the BCS theory, becomes inapplicable\cite{DJ_scalapiono_highSC_BCS,keimer2015_HighSC_BCS}.
		The single-band Hubbard model on the two-dimensional square lattice has been widely used to explore the physics of 
		high-$T_c$ cuprates, and some phenomena observed in experiments have been successfully reproduced~\cite{ZB_Huang2003_2DH,Dagotto1994challenge,Kun_Jiang2018_showSC,keimer2015_HighSC_BCS,
			Lee2006_showSC,PW_Anderson_showSC,DJ_scalapiono_highSC_BCS}. These phenomena include antiferromagnetism at half-filling and a competition of orders at $1/8$ doping~\cite{Antifer_JE_Hirsch1989,Antifer2_SR_White1989,Antifer3_MP_qin2016,Stripes_A_Wietek}.  
		However, numerical analyses revealed that the ground state of the pure $t$-$U$ Hubbard model is not the $d$-wave superconducting 
		(SC) state but instead a stripe phase in which charge density waves (CDW) and spin density waves (SDW) coexist, only partially aligning with 
		experimental findings\cite{CDWSDW_White2003,Stirpe_exp_BX_zheng2017,tUnoSC_Mp2020_qin,Stripe2_ID_Kota2018,tU_C_Halboth2000,tU_FC_zhang1988}.
		
		Theoretical studies have been also performed beyond the pure $t$-$U$ Hubbard model. One example that has been extensively studied is the $t$-$t^\prime$-$U$ Hubbard model, which 
		takes into account the next-nearest-neighbor (NNN) hopping term $t'$. A stripe state with 
		wavelength $\lambda_{c} = 4$ and a quasi-long-range SC order have been reported for the $t$-$t^\prime$-$U$ Hubbard model on a four-leg cylinder~\cite{NNNLamada4exp_Tranquada,NNN_N_Plonka,NNN_Jiang,NNN_Ponsioen,NNNLamada4exp_Tranqueda1,NNNCM_Chung,NNNScience_HC_Jiang,ZSzhou2023}.
		Another example is the two-dimensional checkerboard Hubbard model. In the presence of inhomogeneous hopping integrals, it was shown that 
		such a model harbors $d$-wave superconductivity,
		$d$-wave Mott insulator as well as some other phases~\cite{inhomogeneous_Tsai,inhomogeneous2_Hong_Yao,inhomogeneous3_Karakonstantakis,inhomogeneous4_Baruch,
			inhomogeneous5_Doluweera,inhomogeneous6_Chakraborty,inhomogeneous7_Ying}.
{Recent experiments indicated that there exists anomalously strong nearest-neighbor (NN) attraction $V$ in the one-dimensional cuprate $\mathrm{Ba}_{2+x}\mathrm{Sr}_{x}\mathrm{CuO}_{3+\delta}$~\cite{ZXshen} and in the prototypical cuprate ladder $\mathrm{Sr}_{14}\mathrm{Cu}_{24}\mathrm{O}_{41}$ \cite{PhysRevX.15.021049,NM2025}.
		These findings suggest that $V$ may be a key ingredient in the high-T$_c$ cuprates, which reignites research interests on the extended 
$t$-$U$-$V$ Hubbard model.}
		It was found that repulsive $V$ suppresses the SC correlation and enhances CDW, while attractive $V$ can significantly enhance the SC correlation 
		and suppress CDW~\cite{tUV_C_Peng,tUV_LF_Zhang,tUV_M_Jiang,tUV_R_Micnas}. 
		Above examples demonstrate that the models beyond the pure $t$-$U$ Hubbard model may provide insights into the physics of high-$T_c$ cuprates.        
		
		Recently, a quantum Monte Carlo study of the extended checkerboard Hubbard model indicated that high-temperature $d$-wave superconductivity 
		can be realized via the combination of NN attraction and electron hopping inhomogeneity~\cite{inhomogeneous8_Huang}. 
		Due to the difficulty in controlling error bars at large $U$, quantum Monte Carlo simulations were conducted in the parameter regime of $0\le U \le6t$.
		To gain further insight into the behavior in the strong-coupling regime, we investigate the effects of NN attraction and inhomogeneity on superconductivity in 
		the extended checkerboard Hubbard model on the two-leg ladder by using the density-matrix renormalization group (DMRG)~\cite{DMRG1,DMRG2,DMRG3} method. 
		
		Our results show that, in both the homogeneous and inhomogeneous cases, the intraplaquette NN attraction $V$ enhances the SC correlation. 
		In the homogeneous case, only strong intraplaquette NN attraction (large $|V|$) can induce the SC ground state.
		In the inhomogeneous cases, the critical $|V_{c}|$ required by the formation of SC ground state is 
		greatly reduced with the increase of inhomogeneity. 
		This indicates that inhomogeneity drastically amplifies the effect of $V$ on superconductivity. Moreover, both intraplaquette NN attraction and inhomogeneity weaken spin correlation and 
		single-particle correlation. Interestingly, there exists an essential difference between homogeneous and inhomogeneous cases. 
		{In the former case, 
		the intraplaquette NN attraction $V$ suppresses CDW and the SC correlation is anisotropic.
		While in the latter case, $V$ slightly enhances charge fluctuations, and the hole pair has a $C_4$ symmetry.}
		
		This paper is organized as follows. In section II, we briefly introduce the model and some details of DMRG simulation. 
		In section III, we present the results for the homogeneous and inhomogeneous cases, and analyze the effects of inhomogeneity on the SC, spin, charge, and 
		single-particle properties. In addition, we also discuss the pairing symmetry of the SC state in section III. Finally, a summary is given in section IV.
		
		\section{MODEL AND METHOD}\label{SEC:Model}
\begin{figure}[!ht]
			\centering
			\includegraphics[width=0.95\columnwidth, clip]{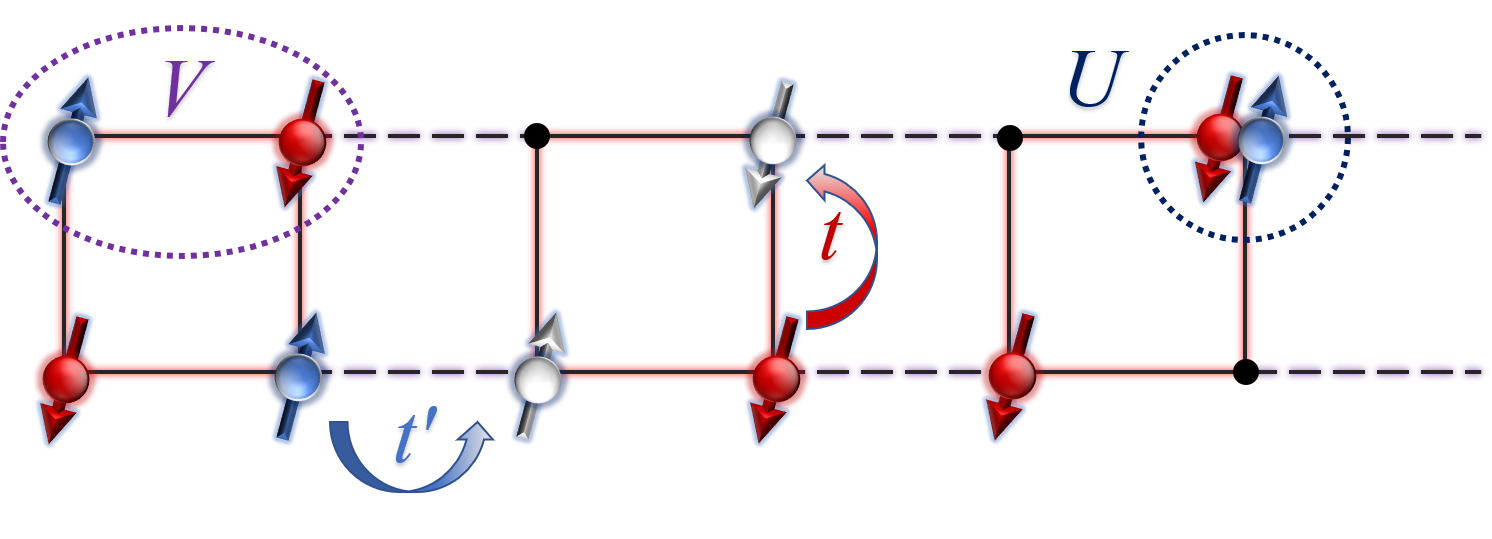}\\
			\caption{A sketch diagram of the checkerboard Hubbard model on the two-leg ladder. $t$ and $t'$ represent NN hoppings within 
				and between plaquettes, respectively. $U$ denotes the on-site repulsion. $V$ stands for the intraplaqutte NN interaction. 
			}\label{Fig01}
		\end{figure}

		The checkerboard lattice consists of periodically arranged $2\times2$ plaquettes, which is illustrate in Fig.~\ref{Fig01}.
		On such a lattice, the Hubbard model on the two-leg ladder is defined as 
		\begin{align}\label{EQ:H}
			\mathcal{H} = 
			&-t \sum_{\left<ij\right>,\sigma} \left(c_{i\sigma}^{\dag}c_{j\sigma}+\mathrm{H.c.}\right) \nonumber 
			-t' \sum_{\left<ij\right>', \sigma} \left(c_{i\sigma}^{\dag}c_{j\sigma}+\mathrm{H.c.}\right) \nonumber \\
			& +U\sum_{i}n_{i\uparrow}n_{i\downarrow} 
			+V\sum_{\left<ij\right>}n_{i}n_{j},
		\end{align}
		where the first and second terms represent the NN hopping integrals of intraplaquette and interplaquette, respectively. 
		$c_{i\sigma}^{\dag}(c_{i\sigma})$ creates (annihilates) an electron at site $i$ with spin $\sigma$.
		$\left<ij\right>$ and $\left<ij\right>'$ denote the intraplaquette and interplaquette NN summations, respectively.
		The third term represents the on-site repulsion for two electrons with different spins.
		$n_{i\sigma}$ is the number operator of electrons for spin $\sigma$ at site $i$ and $n_i=n_{i\uparrow}+n_{i\downarrow}$.
		The last term describes the NN interactions within the plaquette, i.e., there are four NN interactions within one plaquette.
		
		In this work, we use the DMRG method to simulate Eq.~(\ref{EQ:H}), with the code based on the ITensor library\cite{DMRG4}. Open boundary conditions 
		are used in all calculations. Here we focus on the two-leg ladder with the width $L_{y} = 2$  and length $L_{x} = 64$. 
		The electron filling is defined as $\rho = N_e/N$, where $N=L_{x}\times{L_{y}}$ is the total number of lattice sites, and $N_e$ is the total electron number.
		In numerical calculations, unless otherwise specified, we set $t=1$ as the energy unit, and $U$ and $\rho$ are set to be $8$ and $0.8125$. 
		We keep up to $m = 4000 \sim 8000$ optimal states in our calculations, with a typical truncation error of $\epsilon = 10^{-7}$. On a two-leg ladder, such a small 
		truncation error is enough to guarantee the convergence of our results.

		\section{Results and Discussion}
		To clarify the SC property of the checkerboard Hubbard ladder, the key quantities are the singlet pairing-field operators $\Delta_{\rm{r}}^{\dagger}(x)$ and 
			$\Delta_{\rm{l}}^{\dagger}(x)$. $\Delta_{\rm{r}}^{\dagger}(x)$ is defined as 
			\begin{align}\label{EQ:Phi}
				\Delta_{\rm{r}}^{\dagger}(x) = \frac{c_{(x,0),\uparrow}^{\dag}c_{(x,1),\downarrow}^{\dag}-c_{(x,0),\downarrow}^{\dag}c_{(x,1),\uparrow}^{\dag}}{\sqrt{2}},
			\end{align} 
			{Here, the site index $i$ in $c_{i\sigma}$ is replaced by $(x,y)$ with $x$ and $y~(=0,1)$ being the rung index and leg index, respectively.}
			The subscript r means that the pairing is in the rung direction. 
			Following this convention, we can define $\Delta_{\rm{l}}^{\dagger}(x)$ with the pairing bond along the leg direction.
			In particular, $\Delta_{\rm{l}}^{\dagger}(x)$ is only defined within a plaquette in the checkerboard ladder.
			In the one dimensional models, the SC property can then be diagnosed by the pairing correlation functions $\Phi_{\rm{\alpha\beta}}$, which are defined as  
			\begin{align}\label{EQ:Phi}
				\Phi_{\rm{\alpha\beta}}(x-x_0) = \langle\Delta_{\rm{\alpha}}^{\dagger}(x_0)\Delta_{\rm{\beta}}(x)\rangle,
			\end{align}
			where both $\alpha$ and $\beta$ can take $\rm{l,r}$. To minimize the edge effect, we fix $x_0 = L_x/4$ and choose $x_0\le{x}\le{3L_x/4}$, which is far enough from both edges. 
			In the studied parameter regime, the functions $\Phi_{\alpha\beta}$ always decay algebraically and can be well fitted by 
$B_{sc}(x-x_0)^{-K_{sc}}$. $K_{sc}<1$ indicates that the SC correlation dominates in the ground state~\cite{NNNScience_HC_Jiang,KscKc1}.
		
		The charge distribution can be examined by the charge density profile $\langle{n_{x}}\rangle$, which is defined as $\langle{n_{x}}\rangle = \langle{n_{(x,0)}+n_{(x,1)}}\rangle/2$.
		A certain periodic pattern of $\langle{n_x}\rangle$ signifies the development of CDW.
		The $z$ component of spin correlation function is defined as $G_z(x-x_0) = \langle{S_{(x,y)}^zS_{(x_0,y)}^z}\rangle$, 
		and the single-particle correlation is defined as $G_c(x-x_0) = \langle c_{(x,y),\sigma}^\dagger c_{(x_0,y),\sigma}\rangle$.
		The characteristics of spin and single-particle excitations can be diagnosed by the decay 
		behavior of the corresponding correlation functions: an algebraic fit of $G_z$ or $G_c$ in the form of $B_{\alpha}(x-x_0)^{-K_{\alpha}}$ 
indicates a gapless excitation, whereas an exponential fit in the form of $A_{\alpha}e^{-\frac{x-x_0}{\xi_{\alpha}}}$ signifies a gapped excitation.
		\subsection{HOMOGENEOUS CASE $ {t' = t}$}\label{SEC:case1}
		\begin{figure}[!ht]
			\centering
			\includegraphics[width=1\columnwidth, clip]{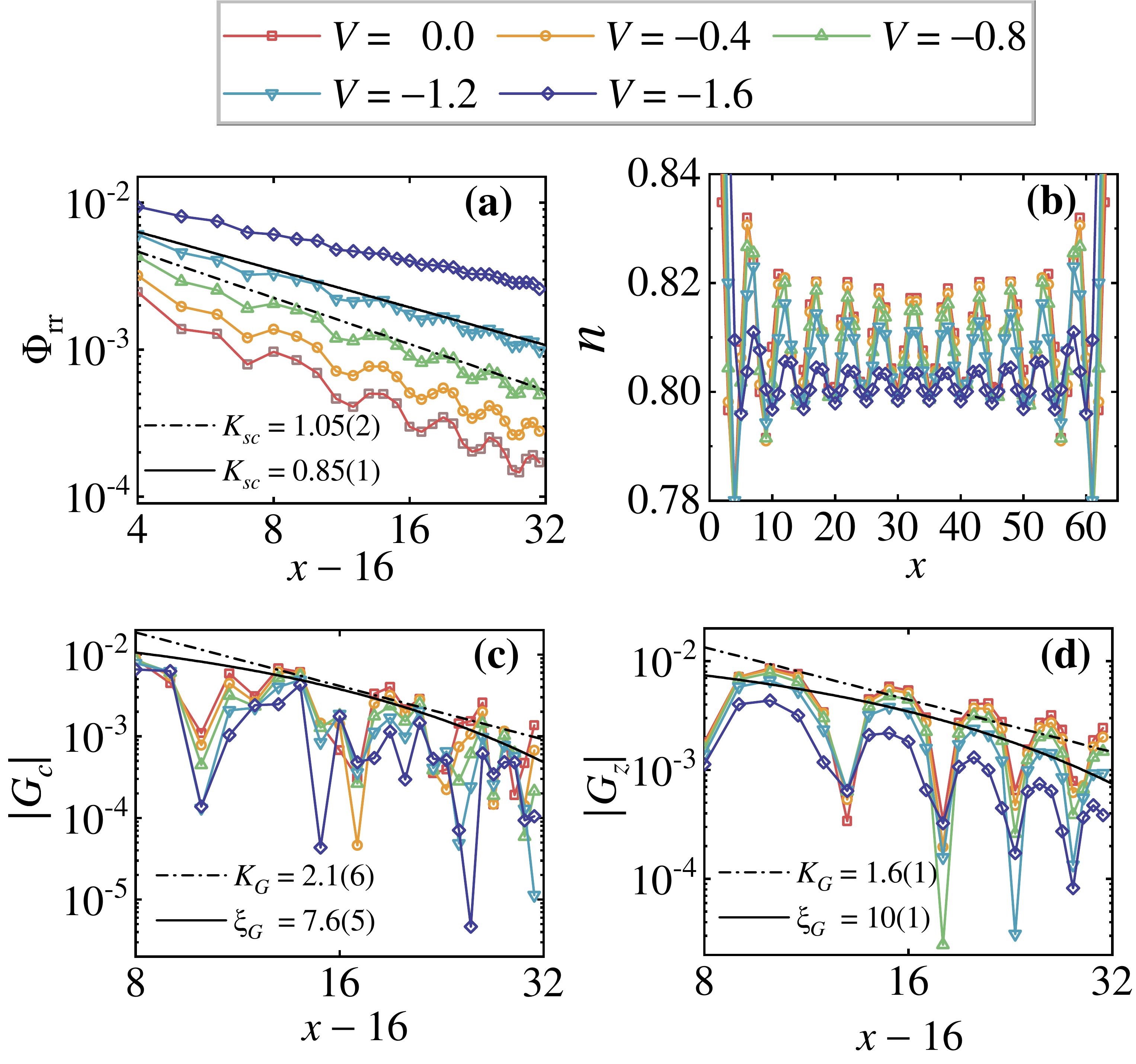}\\
			\caption{ The SC correlation, charge density profile, spin correlation, and single-particle correlation for various 
				intraplaquette NN attractions $V$ in the homogeneous case. 
				(a) shows the singlet pairing correlation function $\Phi_{\rm{rr}}$. The dash-dotted and solid lines show two fitted 
curves of $\Phi_{\rm{rr}}$ at $V = -0.8$ and $V = -1.2$, respectively.
				(b) displays the real-space density profile.
				(c) and (d) show the single-particle and spin correlations. 
				The data at $V = -0.8$ and $V = -1.2$ are well fitted by $B_{\alpha}(x-x_0)^{-K_{\alpha}}$ and  $A_{\alpha}e^{-\frac{x-x_0}{\xi_{\alpha}}}$ respectively, as shown by the dotted and solid lines.
			}\label{Fig02}
		\end{figure}
		
		First, we analyze the effects of the intraplaquette NN attraction $V$ for the homogeneous case $t'=t$.
		To better visualize changes of ordering, we plot the SC, spin and single-particle correlations on logarithmic coordinates. 
		In this coordinate system, algebraic decay manifests as a straight line, while exponential decay as a downward curve.
		In the subsequent analysis, we use $B_{sc}(x-x_0)^{-K_{sc}}$ to fit the two SC correlation curves with $K_{sc} \sim 1$, and show them in
dash-dotted line ($K_{sc} > 1$) and solid line ($K_{sc} < 1$), respectively.
For the single-particle and spin correlations, we use a dash-dotted line to indicate the curve fitted by $B_{\alpha}(x-x_0)^{-K_{\alpha}}$ and a solid line 
to indicate the curve fitted by $A_{\alpha}e^{-\frac{x-x_0}{\xi_{\alpha}}}$.
		
		Fig.~\ref{Fig02} shows $\Phi_{\rm{rr}}$, $n_{x}$, $G_c$ and $G_z$ at $V=0.0, -0.4, -0.8, -1.2$ and $-1.6$.
		From Fig.~\ref{Fig02}(a), it is clear to see that the SC correlation decays algebraically at different $V$ and is enhanced with the increase of 
$|V|$. The fitting of $\Phi_{\rm{rr}}$ indicates that $K_{sc}>1$ when $|V| \leq 0.8$ and $K_{sc}<1$ for $|V|\ge 1.2$, suggesting that the system
		transits to the SC phase at a critical $V_c$ ($0.8<|V_{c}|<1.2$).
Fig.~\ref{Fig02}(b) shows that there exists a weak CDW in the ground state, which is gradually suppressed with increasing $|V|$.
		Figs.~\ref{Fig02}(c) and (d) show that the single-particle and spin correlations are insensitive to $V$ when $|V| \leq 0.8$, but they 
are slightly weakened when $|V|>0.8$. A careful analysis of data reveals that $G_c$ and $G_z$ can be reasonably fitted by algebraical and exponential 
decay formulae when $|V| \leq 0.8$ and $|V| > 0.8$, respectively, implying a transition from gapless to gapped type for the 
single-particle and spin excitations upon entering the SC phase.                 
		The numerical results presented above indicate that in the homogeneous case, $V$ prefers to strengthen the SC correlation and weaken 
		the CDW. This is consistent with the finding in the extended $t$-$U$-$V$ Hubbard model on a four-leg cylinder~\cite{tUV_C_Peng}.
		
		\begin{figure}[!ht]
			\centering
			\includegraphics[width=1\columnwidth, clip]{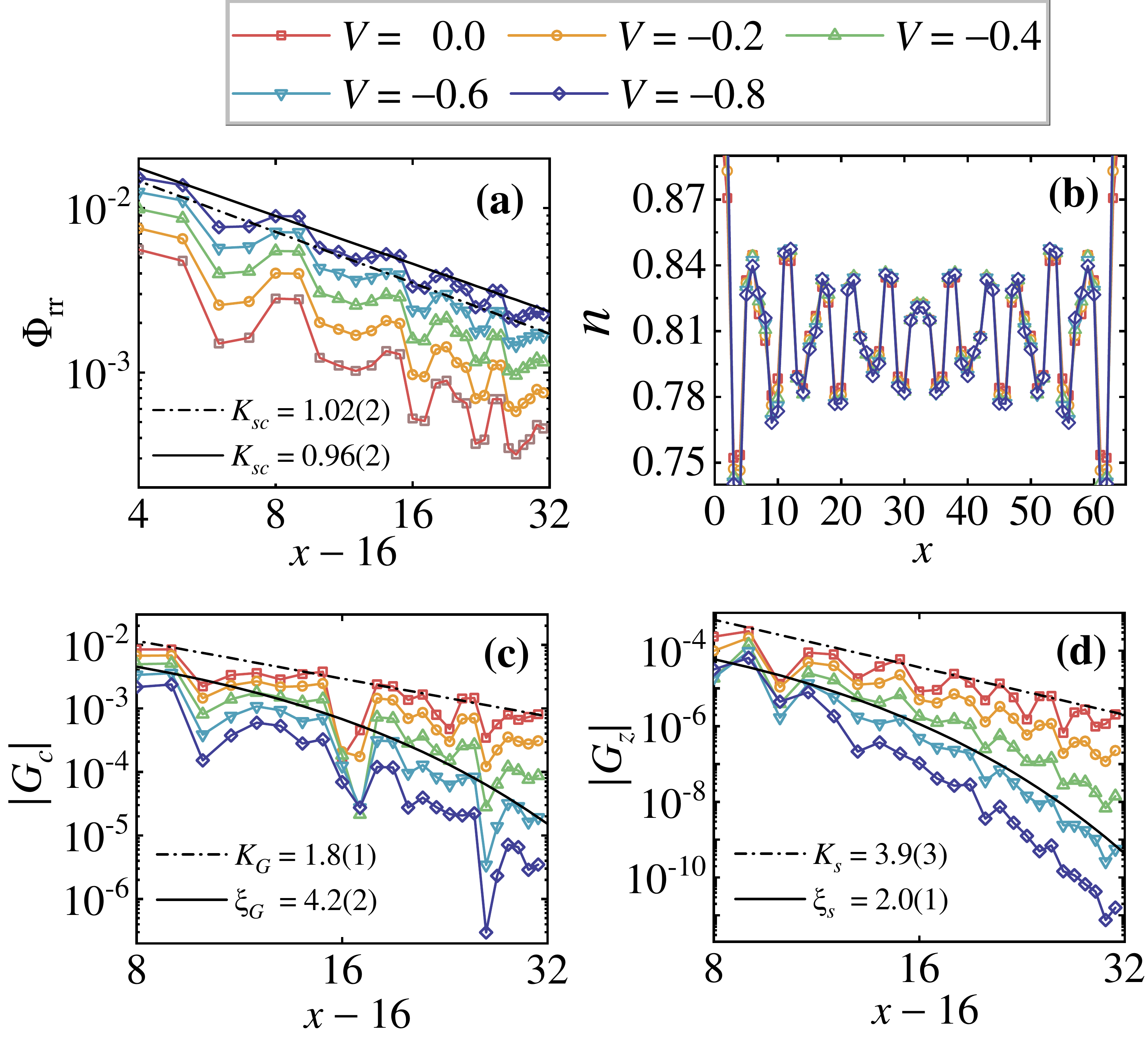}\\
			\caption{Correlation functions and charge density profile at various intraplaquette NN attractions in the inhomogeneous case of $t'=0.4$.
				(a), (c) and (d) show the SC, single-particle and spin correlations, and (b) shows the charge density profile.  
				The algebraical fitting curves for $\Phi_{\rm{rr}}$ at $V=-0.6$ and $V=-0.8$ are plotted in dash-dotted and solid lines, 
and the corresponding $K_{sc}$ values are given in (a). 
				$G_c$ and $G_z$ at $V=0.0$ and $V=-0.6$ are well fitted by $B_{\alpha}(x-x_0)^{K_{\alpha}}$ and $A_{\alpha}e^{-\frac{x-x_0}{\xi_\alpha}}$,
and the corresponding fitting parameters $K_{\alpha}$ and $\xi_{\alpha}$ are given in (c) and (d).
			}\label{Fig03}
		\end{figure}
		
		\subsection{INHOMOGENEOUS CASE $ {t' < t}$}\label{SEC:case2}
\begin{figure}[!ht]
			\centering
			\includegraphics[width=1\columnwidth, clip]{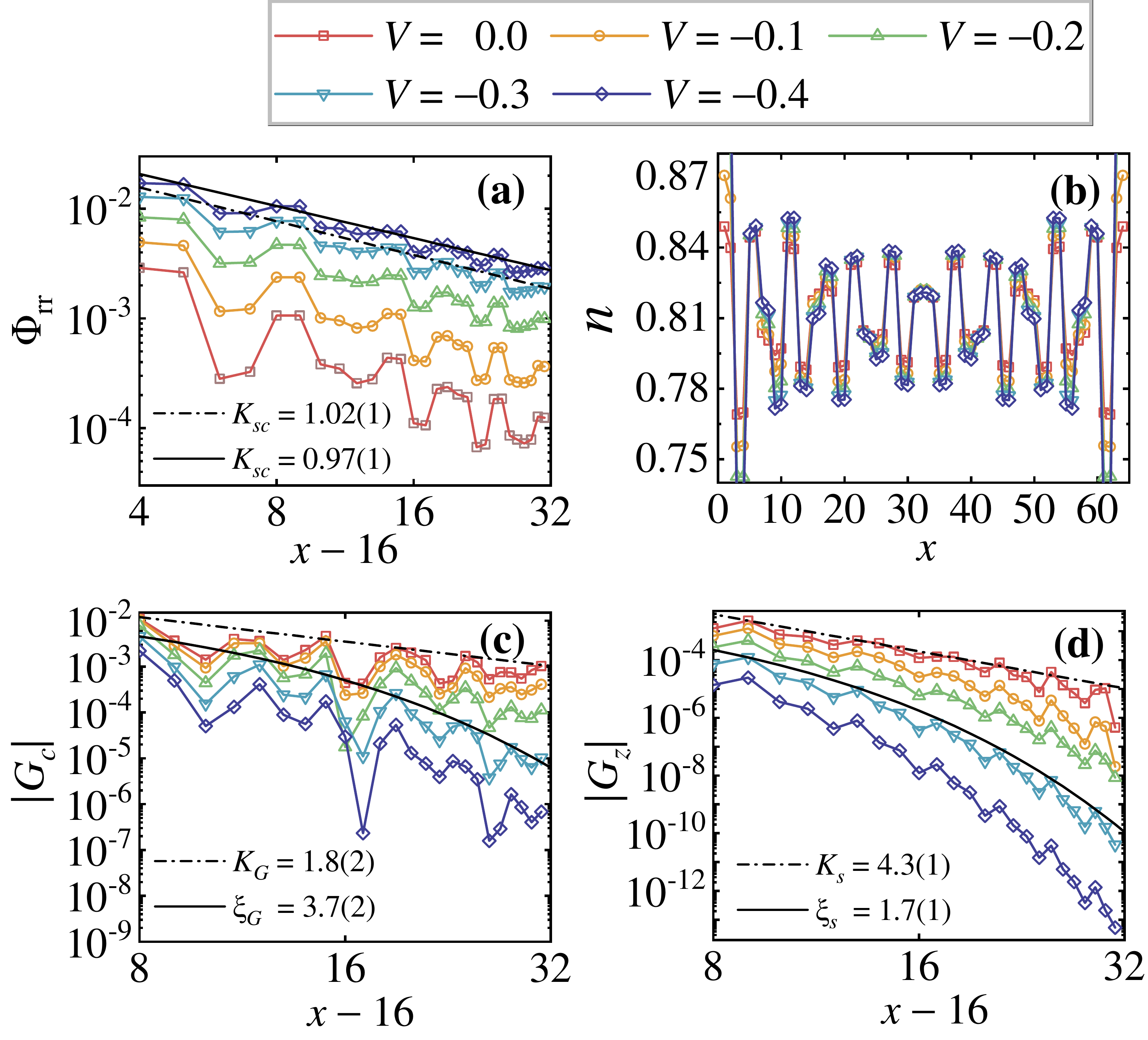}\\
			\caption{Correlation functions and charge density profile at various intraplaquette NN attractions in the inhomogeneous case of $t'=0.1$.
				(a), (c) and (d) show the SC, single-particle and spin correlations, and (b) shows the charge density profile. 
The algebraical fitting curves for $\Phi_{\rm{rr}}$ at $V=-0.3$ and $V=-0.4$ are plotted in dash-dotted and solid lines, 
and the corresponding $K_{sc}$ values are given in (a).  
In (c) and (d), the dash-dotted lines represent algebraic fittings at $V=0.0$, while the solid lines represent power-law fittings at $V=-0.3$.
			}\label{Fig04}
		\end{figure}

		We now turn to analyze the effect of the intraplaquette NN attraction $V$ in inhomogeneous cases. 
		We studied the cases for $t'=0.05$, $0.1$, $0.2$ and $0.4$, and the representative results for $t' =0.4$ and $t'=0.1$
		are shown in Fig.~\ref{Fig03} and Fig.~\ref{Fig04}, respectively.
		Fig.~\ref{Fig03} shows the $V$-dependence of $\Phi_{\rm{rr}}$, $n_{x}$, $G_c(x-x_0)$ and $G_z(x-x_0)$
		for $t'=0.4$, and similar results for $t'=0.1$ are shown in Fig.~\ref{Fig04}. 

        As seen from Figs.~\ref{Fig03}(a), 
        the SC correlation is strengthened with the increase of $|V|$, which is similar to the finding
		for the homogeneous case.
		Interestingly, in the inhomogeneous case, the SC correlation is more sensitive to the 
		intraplaquette NN attraction. At $|V|=0.8$, $K_{sc}<1$ indicates that superconductivity dominates the ground state 
		at a smaller $|V|$ compared to the homogeneous case.
		This demonstrates that inhomogeneity can amplify the effect of intraplaquette NN attraction.
		Unlike the homogeneous case, the charge density profile shown in Fig.~\ref{Fig03}(b) does not exhibit a periodic pattern, implying that 
no CDW is developed at $t' =0.4$. Moreover, increasing $|V|$ leads to stronger inhomogeneous charge distribution. Such a simultaneous enhancement of
SC correlation and charge fluctuation is similar to the effect of $V$ in the two-dimensional $t$-$U$-$V$ Hubbard model, suggesting that the physics in the 
inhomogeneous ladder captures the essential characteristics of the two-dimensional system\cite{CDWenhance2015,CDWenhance2022}. We will explore this in Part C of Section III.
		From Figs.~\ref{Fig03}(c) and \ref{Fig03}(d), it is clear that the single-particle and spin correlations are suppressed by $V$,
		exhibiting obvious algebraical decay at $V=0.0$ and turning to exponential
decay with increasing $|V|$. These results indicate that transiting to the SC phase accompanies with a transition of single-particle and spin excitations
from gapless to gapped type. Notice that the change of excitation is much more clear at $t'=0.4$ than at $t'=1.0$, which also manifests for
smaller $t'$ (see the following figures).
		
		Fig.~\ref{Fig04}(a) shows that with increasing $|V|$ from $0.0$ to $0.4$, the SC correlation is rapidly enhanced by $V$, and 
the system transits to the SC phase at $V=-0.4$. The much smaller $|V_{c}|$ for $t'=0.1$ than the ones for $t'=0.4$
		and $t'=1.0$ demonstrates that strong inhomogeneity favors the formation of superconductivity. Fig.~\ref{Fig04}(b) shows that $V$ play a similar 
role on charge fluctuation to the one at $t'=0.4$.     
		One can readily see from Figs.~\ref{Fig04}(c) and \ref{Fig04}(d) that the single-particle and spin correlations remain algebraical decay 
at $V=0.0$, and transit to exponential decay with increasing $|V|$, indicating a shift from gapless to gapped excitations.
		\begin{figure}[!ht]
			\centering
			\includegraphics[width=1\columnwidth, clip]{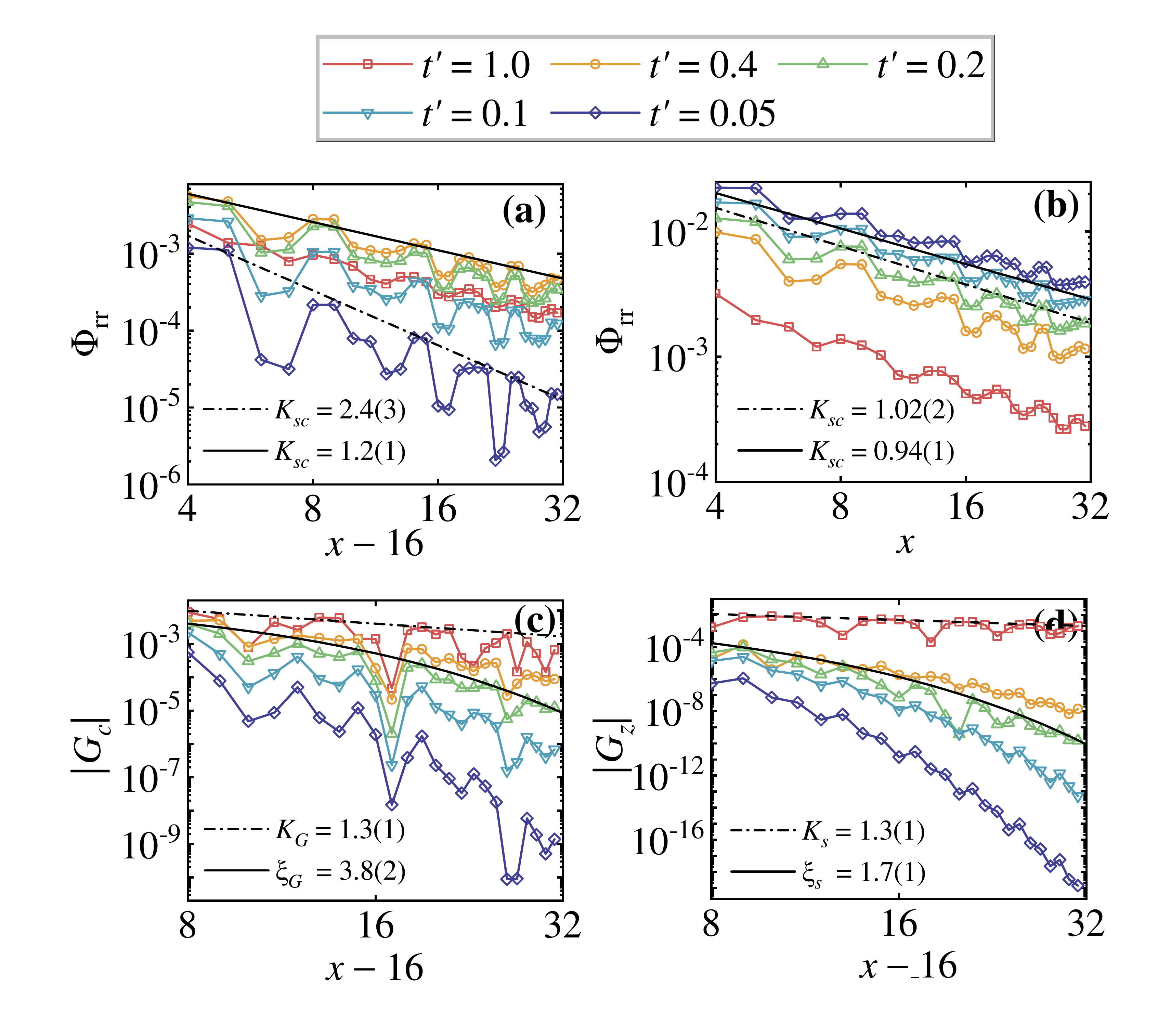}\\
			\caption{Correlation functions at different interplaquette hopping integrals $t' = 1, 0.4, 0.2, 0.1, 0.05$ and various $V$. 
				(a) and (b) show the effect of $t'$ on the SC correlation at $V=0.0$ and $V=-0.4$, respectively.
				(c) and (d) show the effect of $t'$ on the correlation functions of single-particle and spin at $V=-0.4$.
				In (c) and (d), the dash-dotted lines represent algebraic fittings at $t'=1.0$, while the solid lines represent power-law fittings 
at $t'=0.2$.
			}\label{Fig05}
		\end{figure}
		
		For a better understanding of the effect of inhomogeneity, we make a comparison of the results obtained from different
		interplaquette hoppings. Fig.~\ref{Fig05}(a) shows the SC correlations at $V=0.0$ for $t'=1.0, 0.4, 0.2, 0.1$, and $0.05$. 
		It can be seen that the SC correlation exhibits a non-monotonic dependence on $t'$. It increases slightly as 
		$t'$ is reduced from $1.0$ to $0.4$, and then is slightly suppressed as $t'$ is reduced to $0.2$, followed by
		one order of magnitude reduction with further decreasing of $t'$ to $0.05$. 
        {This non-monotonic behavior might have an intimate 
        relation to the optimal pair binding at an intermediate hopping\cite{PhysRevB.77.214502,PhysRevB.96.064527}.}
		As seen from Fig.~\ref{Fig05}(b), at $V=-0.4$, the SC correlation is monotonically increased with the increase of $t'$. 
		For $t'>0.1$, $K_{sc}>1$ indicates that the system lies in the normal state, and when $t'\le 0.1$, $K_{sc}<1$ 
		signifies that the system enters the SC phase.
		Fig.~\ref{Fig05}(b) indicates that at a fixed $V$, increasing inhomogeneity can trigger the appearance of superconductivity.
		Figs.~\ref{Fig05}(c) and \ref{Fig05}(d) show that at $V=-0.4$, a decrease of $t'$ makes the single-particle and spin correlations
		change from algebraic to exponential decay, implying opening a gap in the corresponding excitation.
		
\begin{figure*}[t]
			\centering
			\includegraphics[width=1\textwidth]{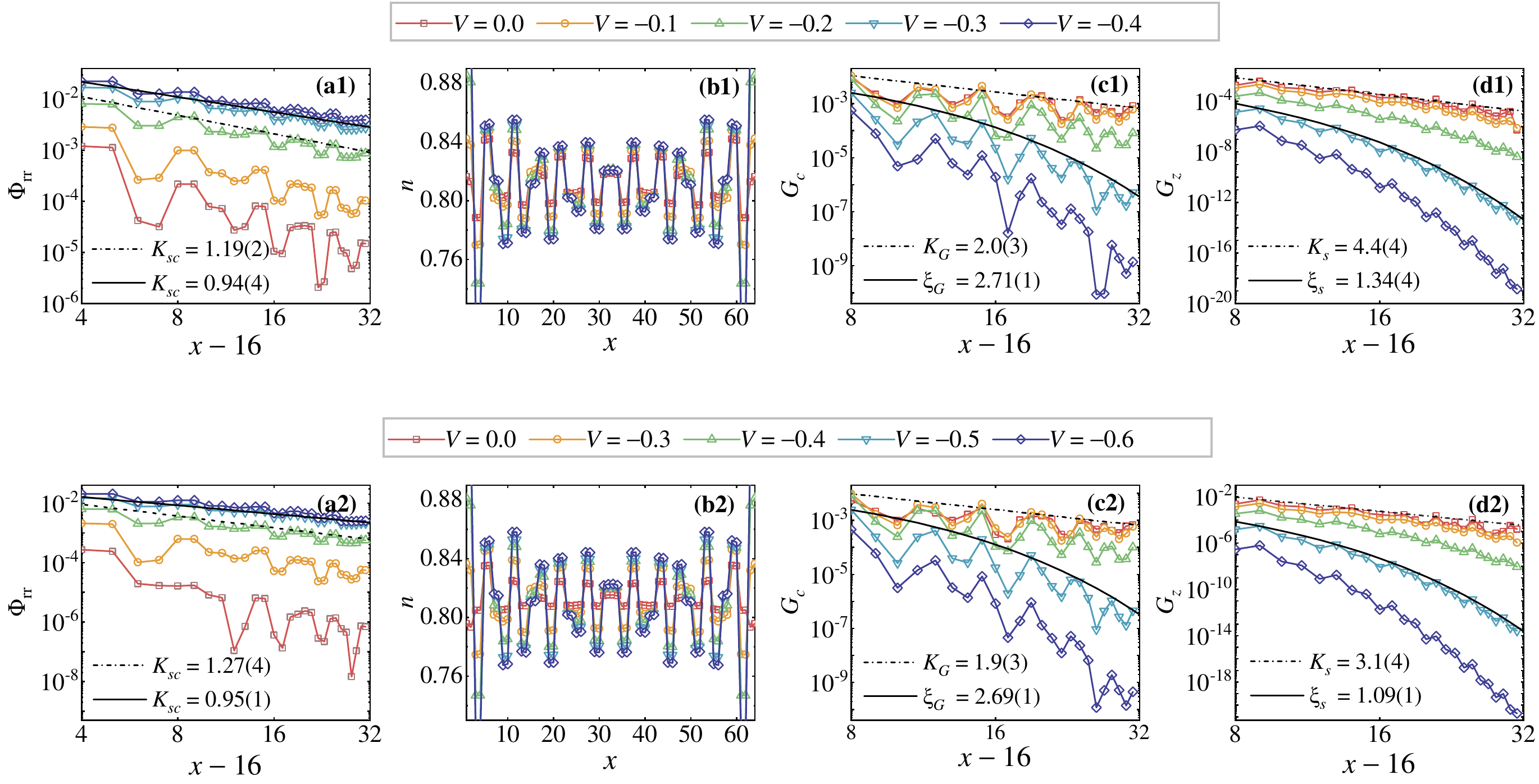}\\
			\caption{(a1)-(d1) Correlation functions and charge density profile at $U=8$ and $t'=0.05$ for $V=0.0, -0.1, -0.2, -0.3$ and $-0.4$.
				(a2)-(d2) Correlation functions and charge density profile at $U=12$ and $t'=0.05$ for $V=0.0, -0.3, -0.4, -0.5$ and $-0.6$.
			}\label{Fig06}
		\end{figure*}

		To understand the effect of $U$ on the $V$-enhanced superconductivity, we make a comparison for the results at different $U$.
        In Figs.~\ref{Fig06}(a1)-\ref{Fig06}(d1), we present the results for $U=8$ and $t'=0.05$ at different $V$. The fitting of $\Phi_{\rm{rr}}$ 
        shows that the system enters the SC phase when $V=-0.3$, and the SC correlation is larger than that at $t'=0.1$ and $V=-0.4$, 
		as shown in Fig.~\ref{Fig06}(a1). A combination of the above results reveals that for the fixed $U=8$,          
		stronger inhomogeneity indicates stronger $V$-induced SC enhancement effect. This is clearly manifested by 
		a rapid decrease of the critical $V_{c}$ with decreasing $t'$. 
		
		In Figs.~\ref{Fig06}(a2)-(d2) we present the simulation results for $U=12$ and $t'=0.05$. 
		Comparing Fig.~\ref{Fig06}(a1) and \ref{Fig06}(a2), one can see that at $V=0.0$, the magnitude of SC correlation for $U=12$ is one order smaller 
		that the one for $U=8$, which can be attributed to the decrease of quasiparticle weight with increasing the on-site interaction.  
		In the case of $U=8$, $K_{sc}<1$ when $V=-0.3$ and $K_{sc}>1$ when $V=-0.2$, suggesting that the critical $V_c$ for the formation
		of superconductivity is between $-0.2$ and $-0.3$.
		In the case of $U=12$, $K_{sc}<1$ when $V=-0.5$ and $K_{sc}>1$ when $V=-0.4$, indicating that the critical $V_c$ lies 
		between $-0.4$ and $-0.5$. 
		We can see that the amplitude of the critical $V_c$ increases with the increase of $U$.
		A comparison between Fig.~\ref{Fig06}(b1) and \ref{Fig06}(b2) shows that charge distribution is similar away from the ladder edges for 
		$U=8$ and $U=12$. From Figs.~\ref{Fig06}(c2) and \ref{Fig06}(d2), it is clearly seen that $G_z$ and $G_c$ at $U=12$ are enhanced compared with the ones 
		at $U=8$. In particular, the spin and single-particle excitations for $U=12$ also exhibit a transition from gapless to gapped characteristic upon entering the SC phase.  
		
		Finally, we calculated the extended checkerboard Hubbard model for hole doping concentrations $\delta=0.5$ and $\delta=0.0$ 
		at $t'=0.05$ and $V = -0.4$ (the results not shown here). At $\delta=0.0$, the system lies in an insulating antiferromagnetic state. 
		At $\delta=0.5$, the ground state is a hardcore boson insulating state in which each plaquette 
		contains a pair of holes~\cite{inhomogeneous3_Karakonstantakis,hardcoreboson}.
		
		In order to understand the $V$-enhanced superconductivity, we carried out exact diagonalization~(ED) calculations
		for an isolated $2\times2$ plaquette, and the obtained pair binding energy and clustering energy are listed in Table~\ref{tab1}.
		The pair binding energy is defined as 
		\begin{align}\label{EQ:Phi}
			E_b = E(2,2) + E(1,1) - 2E(2,1),
		\end{align} 
		where $E(n_1,n_2)$ is the ground energy of the isolated plaquette with $n_1$ spin-up and $n_2$ spin-down electrons. 
		$E_b<0$ indicates that the plaquette favors hole pairing.
		The clustering energy is written as
		\begin{align}\label{EQ:Phi}
			E_c = E(3,3) + E(1,1) - 2E(2,2).
		\end{align} 
		$E_c < 0$ means that half-filled plaquette tends to separate into two hole-rich phase and two electron-rich phase~\cite{PSstate}.
		
		\begin{table}[b]
			\centering
			\caption{The ED results for the isolated $2\times2$ plaquette at $U = 8$ and $U = 12$.
				The change of pair binding energy from positive to negative indicates the formation of hole pairs.}
			\label{tab1}
			\begin{ruledtabular}
				\begin{tabular}{ccc}
					& $U = 8$ $V = -0.20$ & $U = 8$ $V = -0.22$ \\
					\colrule
					$E_b$  & 0.00267991 & -0.008780338 \\
					$E_c$  & 0.27024210 & 0.2584706757 \\
				\end{tabular}
				\begin{tabular}{ccc}
					& $U = 12$ $V = -0.42$ & $U = 12$ $V = -0.44$ \\
					\colrule
					$E_b$  & 0.00728448 & -0.0042018833 \\
					$E_c$  & 0.33580341 & 0.3239385558 \\
				\end{tabular}
			\end{ruledtabular}
		\end{table}

		Table~\ref{tab1} shows that $E_b$ changes from a positive value to a negative value at $V\sim-0.2$ for $U=8$ and at $V\sim-0.42$ for $U=12$,
		respectively. This indicates that the hole pairs are formed in plaquettes when $V<-0.2$ for $U=8$ and $V<-0.42$ for $U=12$. 
		The positive $E_c$ in Table~\ref{tab1} can safely exclude the suppression of phase separation on superconductivity. 
        The good agreement between the transition $V_c$ estimated from DMRG and ED at $t'=0.05$ demonstrates that the formation of hole 
		pairs is crucial for the emergence of off-diagonal SC order.
        An increasing of $t'$ benefits the phase coherence between hole pairs~\cite{inhomogeneous_Tsai,inhomogeneous2_Hong_Yao}, 
		and meanwhile, it is harmful for the stability of hole pairs. The rapid increase of $|V_c|$ with increasing $t'$ evidences that the 
		harmful effect is dominant and stronger $|V|$ is required to stabilize hole pairs when the $t^\prime$ becomes larger.   
		In the weak- and intermediate-coupling regimes ($0\le U\le4$), quantum Monte Carlo simulations also showed that at $t'=0.05$, 
		the SC phase is established when $V<0.0$, wherein $E_b$ is negative~\cite{inhomogeneous8_Huang}.

		\subsection{PAIRING SYMMETRY}\label{SEC:Pairing symmetry}
		Finally, we briefly discuss the pairing symmetry related to the superconductivity. Physically, the two-legged ladder does not have the same spatial symmetry along the leg and rung directions, but it can still give us some insights into the pairing symmetry of the two-dimensional extended checkerboard Hubbard model. Here, three different SC correlations, 
		$\Phi_{\rm{rr}}$, $\Phi_{\rm{ll}}$, $-\Phi_{\rm{lr}}$, are used to judge the symmetry.
		\begin{figure}[!ht]
			\centering
			\includegraphics[width=1\columnwidth]{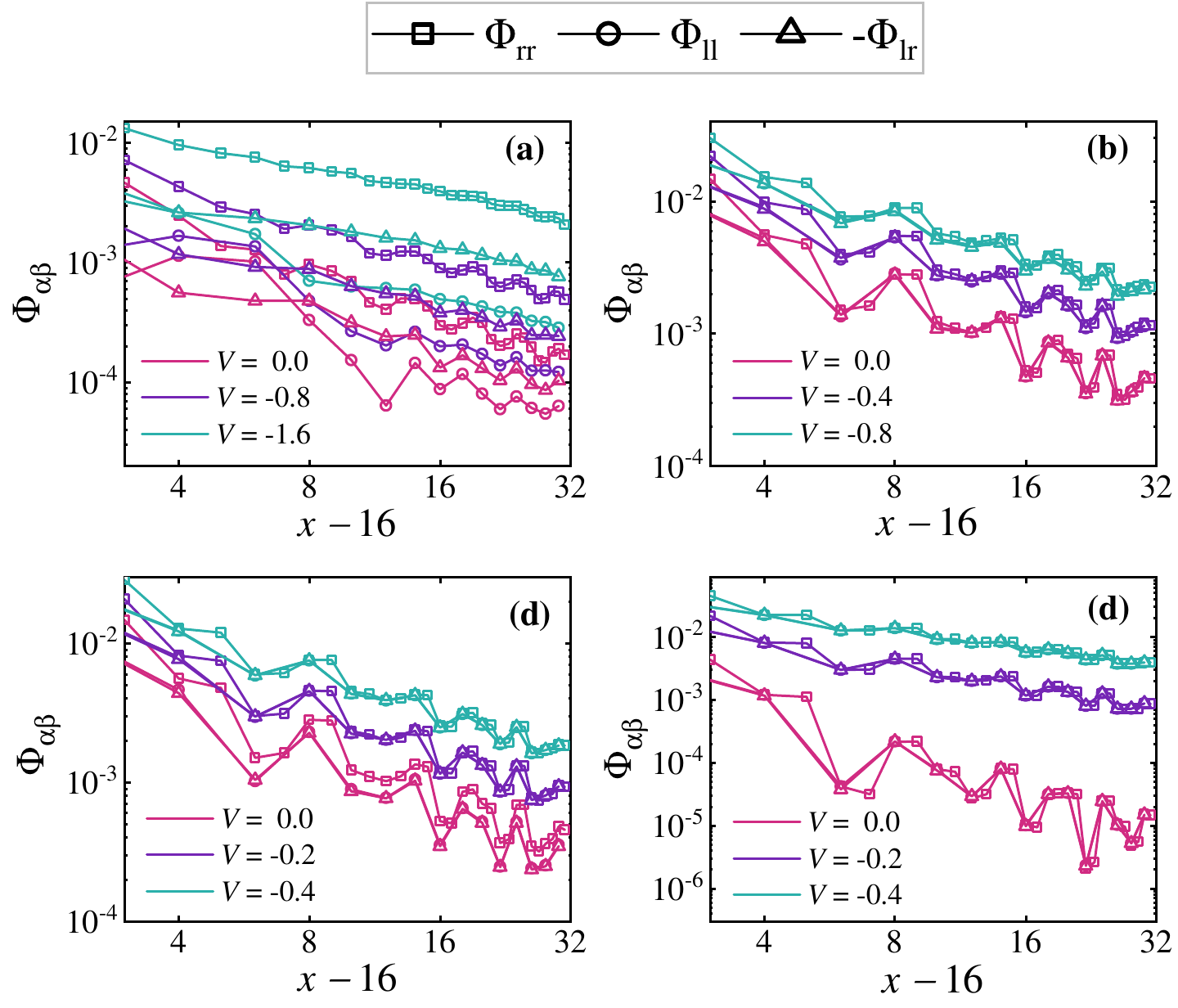}\\
			\caption{Three SC correlations $\Phi_{\rm{rr}}$, $\Phi_{\rm{ll}}$, and $-\Phi_{\rm{lr}}$ are shown for different intraplaquette NN attractions. 
				(a) shows the homogeneous case with $t'=1.0$. (b), (c) and (d) show the inhomogeneous cases with $t'=0.4, 0.2$ and $0.05$, respectively.
			}\label{Fig07}
		\end{figure}
		
		Fig.~\ref{Fig07} shows the SC correlations for different $t'$ and $V$. 
		Firstly, we can find that both $\Phi_{\rm{rr}}$ and $\Phi_{\rm{ll}}$ are positive, but $\Phi_{\rm{lr}}$ is negative. This is the characteristic of 
		d-wave pairing symmetry. 
		Secondly, $\Phi_{\rm{rr}}$, $\Phi_{\rm{ll}}$ and $-\Phi_{\rm{lr}}$ are significantly enhanced by $V$ and the reduction of $t'$ drastically 
		intensifies the effect of $V$ on superconductivity. Interestingly, there exists a qualitative difference between the homogeneous and  
		inhomogeneous cases. Fig.~\ref{Fig07}(a) shows that in the homogeneous case, $\Phi_{\rm{rr}}$, $\Phi_{\rm{ll}}$ and $-\Phi_{\rm{lr}}$ are completely 
		different. The magnitude of $\Phi_{\rm{rr}}$ is about one order larger than that of $\Phi_{\rm{ll}}$ and $-\Phi_{\rm{lr}}$. On the other hand, 
		Fig.~\ref{Fig07}(b) shows that $\Phi_{\rm{rr}}$, $\Phi_{\rm{ll}}$ and $-\Phi_{\rm{lr}}$ are almost indistinguishable at $t'=0.4$, suggesting that 
		the hole pair has a $C_4$ symmetry.
		Figs.~\ref{Fig07}(c) and \ref{Fig07}(d) show the results for $t'=0.2$ and $0.05$, respectively. The behaviors of the SC correlations are very similar to that for $t'=0.4$. 
		Therefore, in the inhomogeneous cases, it can be regarded that the two-leg-ladder Hubbard model captures the physics of the two-dimensional checkerboard Hubbard model.

		\section{CONCLUSIONS}\label{SEC:conclusions}
		
		In summary, we have systematically investigated the effect of inhomogeneity on the ground state of the extended checkerboard 
		Hubbard model on a two-leg ladder.
		Our DMRG results show that in the inhomogeneous cases, the intraplaquette attraction $V$ dramatically enhances the SC correlation, and
		the enhancement effect becomes stronger as the inhomogeneity increases.
		$|V_c|$ required for the formation of superconductivity is reduced from $1.2$ at $t'=1.0$ to $~0.3$ at $t'=0.05$.
		Whatever the homogeneous or inhomogeneous case, both the single-particle and spin excitations open a gap in the SC phase.
		One significant difference between homogeneous and inhomogeneous cases is that while the hole pairing is asymmetric along 
		the rung and leg directions in the former case, the $C_4$ symmetry inherent for the two-dimensional lattice is applicable 
		for the latter case. A combination of DMRG and ED results reveals that in the extremely inhomogeneous case, the SC phase is
		established after the formation of hole pairs in plaquettes. Our numerical results confirm that in the strong-coupling regime,
		which is the physically relevant, the combination of electronic inhomogeneity and NN attraction can indeed promote the formation
		of $d$-wave superconductivity. 
		
		\begin{acknowledgments}
			We acknowledge funding support from the Ministry of Science and Technology of the People's Republic of China (Grant No.~2022YFA1402704), the National Natural Science Foundation of China (Grants Nos.~12274187, and 12247101) and the Fundamental Research Funds for the Central Universities (Grant No.~ lzujbky-2024-jdzx06), the Natural Science Foundation of Gansu Province (No. 22JR5RA389).
		\end{acknowledgments}
		
		\bibliographystyle{apsrev4-2}
		\bibliography{ref.bib}

	\end{CJK*}
	
	
\end{document}